\documentclass[11pt]{article}
\usepackage[margin=2.45cm]{geometry}
\usepackage{mathptmx}
\usepackage{amsmath,amssymb,amsthm}
\usepackage{graphicx}
\usepackage{booktabs}
\usepackage{algorithm}
\usepackage{algpseudocode}
\usepackage{float}
\usepackage{caption}
\usepackage[hidelinks]{hyperref}
\usepackage{natbib}
\newtheorem{theorem}{Theorem}
\newtheorem{lemma}{Lemma}
\newtheorem{proposition}{Proposition}
\newcommand{\R}{\mathbb{R}}
\newcommand{\cG}{\mathcal{G}}
\newcommand{\cT}{\mathcal{T}}
\title{\bfseries A Global Solver for Minimum-Time Connectivity Restoration in UAV Networks}
\author{Huan Lin, Lianghui Ding\thanks{Corresponding author: Lianghui Ding (lhding@sjtu.edu.cn).}\\[4pt]
\small Institute of Image Communication and Network Engineering, School of Integrated Circuits,\\
\small Shanghai Jiao Tong University, Shanghai 200240, China\\
\small \texttt{lhzt715@sjtu.edu.cn}; \texttt{lhding@sjtu.edu.cn}}
\date{}
\begin{document}
\maketitle
\begin{abstract}
Node failures can split an unmanned aerial vehicle (UAV) network into disconnected groups. Restoring communication requires moving the surviving aircraft, and the operator needs both an executable movement plan and an assessment of how much shorter the recovery could be. We develop Connectivity Restoration Branch-and-Bound (CR-BB), which returns a feasible plan and numerical displacement bounds; dividing these bounds by the common speed gives the corresponding time interval. Exact-arithmetic guarantees are established by the theorems. When the gap is not closed, the plan is feasible but not certified optimal. A fixed-tree subproblem is a second-order cone program. CR-BB searches over spanning trees and uses a Joint Mandatory-Link Bound (JMLB) to remove tree families that cannot improve the incumbent; Progress-based Adaptive Stopping (PAS) controls the search budget. On 200 matched cases, enabling the JMLB in the branch bound raised the number of cases meeting the prescribed displacement gap from 182 to 200 and reduced mean expanded nodes from 49.6 to 0.7. On a 60-case exact comparison, CR-BB met the gap in 90.0\% of cases, compared with 55.0\% for a spanning-tree mixed-integer conic formulation and 37.5\% of the eight runs completed by an independent optimizer. On 1,000 paired recovery cases, CR-BB planned first restored connectivity in 72.35 steps on average, compared with 124.32 for an edge-augmentation baseline with conic refinement, a 41.8\% reduction; on the common 50\% subset the reduction was 28.1\%. Across 2,466 cases, PAS reduced cumulative solver time by 20.9\% and expanded nodes by 31.1\% relative to a fixed search budget, with a 0.36-step mean increase in planned steps; most of the saving came from cases where neither configuration met the gap, and 60 plans became longer.
\end{abstract}

\section{Introduction}
Node failures can leave the surviving UAVs in several disconnected groups. Communication is restored by moving them to new positions, but a plan alone is not enough: the operator also needs to know how much shorter the recovery could be. A plan without a bound may be far from the best possible; a bound without a plan is not executable. We therefore ask for both at once: what is the earliest terminal recovery, what movement realizes it, and what proves that no earlier recovery exists?

The survivors start at known positions and move simultaneously at a common maximum speed. Two aircraft communicate when their distance is at most $R$. A plan assigns a final position to every survivor, and its cost is the largest distance traveled by any aircraft. If each aircraft can move at most $s$ metres per step, a plan of cost $D$ is executed in at most $\lceil D/s\rceil$ steps by straight-line interpolation. The largest displacement, not the sum of displacements, therefore determines the recovery time.

The optimization problem is to choose a connected terminal configuration that minimizes this largest displacement. For a fixed spanning tree, the best terminal positions solve a second-order cone program: each aircraft stays within a trial radius of its start, and each tree edge is at most $R$. The difficulty is the choice of tree. We solve it with a best-first branch-and-bound search. At a trial displacement, a pair can become an edge only if its initial separation is at most $R+2U$. Bridges of the resulting candidate graph are mandatory in every improving tree. For a set of mandatory links, a conic dual gives a lower bound on the displacement required by every tree in that branch. The search returns a verified terminal configuration, a valid lower bound, and a stopping status; if a node or time limit interrupts it, both bounds remain valid.

The technical contributions are as follows.
\begin{enumerate}
\item \textbf{A global solver with two outputs.} We call the solver \emph{Connectivity Restoration Branch-and-Bound} (CR-BB). It returns terminal positions, a recovery plan, and displacement bounds $[L,U]$; the corresponding time interval is $[L/v,U/v]$. The fixed-tree SOCP gives the positions for a fixed tree; the exact-arithmetic interval guarantee comes from searching over connected terminal configurations and retaining the bounds when the search stops.
\item \textbf{A joint lower bound from shared-relay geometry.} We call this the \emph{Joint Mandatory-Link Bound} (JMLB). The subproblem is a standard SOCP dual; the new element is identifying the mandatory edges and aggregating their shared-endpoint directions into one branch bound. A relay shared by several mandatory links cannot occupy different terminal positions for different links. Evaluating the incident directions together gives a lower bound for a whole set of trees, so one comparison can remove that set. We prove that the bound applies to every tree in the branch, that the branches cover every tree that could improve the current plan, and that $L\le D^\star\le U$ at any stop. On 200 matched 50\% cases, branch JMLB adds 18 gap attainments and reduces mean expanded nodes from 49.6 to 0.7.
\end{enumerate}

PAS is a budget-control policy: it decides when another budget stage is worthwhile, never removes a tree set, and retains active and retired bounds. Section~\ref{sec:pas-method} defines it; Section~\ref{sec:pas-results} reports the resulting time--quality trade-off.

\section{Related Work}
\subsection{Optimization formulations}
Movement minimization asks how far a set of points must move to acquire a prescribed graph property. Demaine et al.~\citep{demaine2009movement} introduced the general framework and its computational difficulty; Anari et al.~\citep{anari2016euclidean} studied the Euclidean objective; and Engin and Isler~\citep{engin2018minimizing} specialized the problem to connectivity of randomly deployed robots. For connectivity under a shared speed limit, the largest displacement is the appropriate objective because the aircraft with the farthest to travel determines the completion time.

Geometric feasibility under uncertainty is closely related. Connectivity graphs of uncertainty regions describe when one position per point can be connected~\citep{chambers2016connectivity}; minimum-spanning-tree problems with neighborhoods optimize a tree under geometric choices~\citep{blanco2017mst}; and classical optimal-tree formulations separate edge-length, bottleneck, and movement objectives~\citep{magnanti1995optimal}. At a fixed displacement limit $U$, each aircraft may choose a position in a ball of radius $U$ around its start, and the question is whether one choice per aircraft makes the communication graph connected. Our candidate graph is the corresponding budget-indexed graph, while the fixed-tree subproblem keeps the original coordinates rather than discretizing a neighborhood.

The continuous subproblem is a second-order cone program; Lobo et al.~\citep{lobo1998socp} and Boyd and Vandenberghe~\citep{boyd2004convex} provide the conic duality used for the lower bounds. The combinatorial layer draws on branch and bound~\citep{land1960automatic}, bridge finding~\citep{tarjan1974note}, and mixed-integer formulation techniques~\citep{vielma2015mixed}; SCIP~\citep{scip2021} and Clarabel~\citep{clarabel2024} are the solvers. Shi et al.~\citep{shi2025fast} give the closest exact formulation: they jointly optimize robot positions and communication links for $k$-connectivity, and at $k=1$ with zero minimum separation their objective and connectivity requirement match ours. Their heuristic first constructs a set of links and then refines the positions; it is a recovery heuristic, whereas Shi OPT is the mixed-integer model. Shi OPT is exact and can certify optimality when it solves. Our comparison at a fixed budget instead reports a displacement interval that remains valid when the solver stops; this guarantee is the object of the present study. The fixed-tree SOCP is exact; the global bounds require the tree search and JMLB below.

\subsection{Resilient recovery}
Restoration planners differ in which links they preserve as the surviving vehicles move. Sensor-network methods reconnect components with a minimum Steiner tree after simultaneous failures~\citep{lee2010recovery}, minimize topology changes after a single failure~\citep{abbasi2013recovering}, optimize the restored partition~\citep{senel2011optimized}, assign channels in centralized restoration~\citep{chouikhi2017centralized}, or manage topology under failures more generally~\citep{younis2014topology}. These methods focus on the cost or structure of the restored network; most do not optimize continuous final positions or provide a lower bound on the earliest recovery.

UAV-specific work spans several execution models. HERO combines hybrid multi-agent restoration~\citep{mi2011hero}, SIDR uses swarm intelligence~\citep{chen2020sidr}, and graph-convolutional methods learn a resilient controller or a scalable representation~\citep{mou2022resilient,lin2024multihop}. Hybrid multi-agent deep deterministic policy gradient (MADDPG) with an artificial potential field (APF) combines reinforcement learning with artificial potentials~\citep{li2025fast}, mission-specific recovery targets electronic-signal tasks~\citep{liu2025robust}, and R2C combines graph learning with centralized fragment planning and periodic replanning~\citep{peng2026resilient}. Coverage can also matter: R2C includes the area served by the recovering swarm, which can change where aircraft should move even when several configurations restore communication equally soon. These planners answer how to recover under their own information models; this paper instead treats the terminal configuration as the decision object and returns a certified distance interval. We therefore use R2C and the Shi edge-augmentation/SOCP refinement as published reference planners.

\section{Problem Formulation}
\subsection{System, damage, and communication}
Let $\mathcal{V}=\{1,\dots,N\}$ be the pre-damage identities, $\mathcal{D}\subset\mathcal{V}$ the destroyed set, and $\mathcal{S}=\mathcal{V}\setminus\mathcal{D}$ the survivors, with $n=|\mathcal{S}|$. The recovery clock starts at $t=0$ with survivor positions $p_i\in\R^d$; destroyed nodes are removed from the communication graph. A configuration $X=\{x_i\}_{i\in\mathcal{S}}$ induces the closed-ball graph $\cG_R(X)$ with an undirected edge $\{i,j\}$ whenever $\|x_i-x_j\|_2\le R$, including equality. Connectivity means that every survivor is in one component, not that every pair has a direct link. If $n=1$ or the initial graph is connected, the answer is zero.

\subsection{Movement and computation time}
A continuous trajectory $q_i(t)$ changes at speed at most $v$. In the discrete execution, $q_i[k+1]=q_i[k]+u_i[k]$ with $\|u_i[k]\|_2\le s$; hovering and a short final step are allowed. If a physical sampling period $\Delta t$ exists, $s=v\Delta t$; otherwise we report completed moves and no physical seconds are inferred. The model is obstacle-free and does not impose collision avoidance or a minimum terminal separation.

\subsection{First connectivity and its terminal equivalent}
The first connectivity time is $T_{\rm fc}=\inf\{t:\cG_R(q(t))\text{ connected}\}$. For any feasible motion of duration $T_{\rm fc}$, every node is within $vT_{\rm fc}$ of its start, so $D^\star\le vT_{\rm fc}$ where
\begin{equation}\label{eq:problem}
D^\star=\min_{X}\ \max_{i\in\mathcal{S}}\|x_i-p_i\|_2
\quad\text{s.t.}\quad \cG_R(X)\text{ connected}.
\end{equation}
Conversely, a terminal configuration at distance $D^\star$ is reached by synchronous straight-line interpolation, giving $T^\star=D^\star/v$. With per-step displacement $s$, the integer optimum is $K^\star=\lceil D^\star/s\rceil$: any $K$-step plan has $D^\star\le Ks$, and a $K^\star$-step linear interpolation attains the terminal configuration. The minimum in \eqref{eq:problem} exists. A common terminal point gives a finite upper bound, so it is enough to search the compact ball of configurations within that bound; the feasible set is a finite union of closed tree-constrained sets, and the minimum of a continuous function on this union is attained. The model is static and centralized; connectivity at the final positions is the only requirement.

\subsection{Tree structure and the computational object}
A connected graph contains a spanning tree, and a feasible tree establishes connectivity. Therefore
\begin{equation}\label{eq:tree}
\min_{T\in\cT_n}\ f(T),\qquad
f(T)=\min_{X,\tau}\ \tau
\ \text{s.t.}\ \|x_i-p_i\|_2\le\tau,\ \|x_i-x_j\|_2\le R\ \forall (i,j)\in T,
\end{equation}
where $\cT_n$ is the set of labeled spanning trees. For fixed $T$, \eqref{eq:tree} is a second-order cone program: the node constraints are cones on $(\tau,x_i-p_i)$ and the edge constraints are cones on $(R,x_i-x_j)$. The global problem is a minimum over finitely many convex problems. A complete graph has $n^{n-2}$ labeled trees, and the decision version is NP-hard; the algorithmic question is how to cover the relevant families without enumeration.

The solver returns a terminal configuration $X$, a spanning tree $T$ whose edges are checked independently, a lower bound $L$, an upper bound $U$ with $L\le D^\star\le U$, and a stopping reason. The bounds $L,U$ are displacements. With common speed $v$, the time interval $[L/v,U/v]$ contains the minimum recovery time $T^\star=D^\star/v$. If a node or time limit stops the search first, the displacement bounds and the derived time interval remain valid; we report the absolute displacement gap $U-L$. A heuristic can supply a plan but not a lower bound.

\section{CR-BB: Exact Search with Guarantees}
\subsection{Overview}
\begin{figure}[t]
\centering
\includegraphics[width=.98\linewidth]{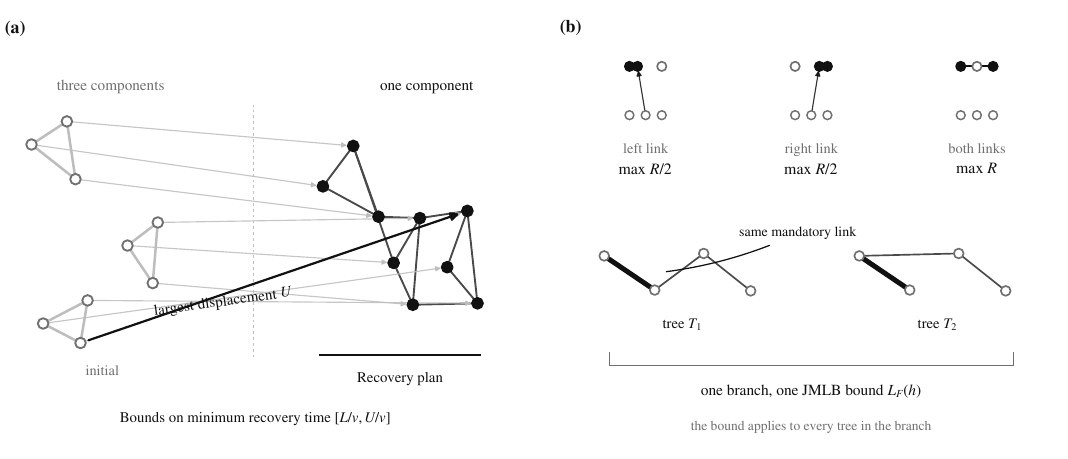}
\caption{Overview of CR-BB. (a) Three disconnected survivor components (left) and the connected terminal configuration returned by the solver (right). Arrows show per-aircraft moves; the highlighted arrow has length $U$. The solver returns a recovery plan and displacement bounds $[L,U]$, equivalently time bounds $[L/v,U/v]$. (b) A relay shared by two mandatory links. Satisfying either link alone needs maximum displacement $R/2$, by moving the relay toward that neighbor. The relay has one terminal position when both links are required, so the joint maximum is $R$. The two bottom trees have different topologies but contain the same mandatory link; one JMLB bound covers the whole branch.}
\label{fig:shared}
\end{figure}

We call the search \emph{Connectivity Restoration Branch-and-Bound} (CR-BB). It reduces the number of tree sets in four steps: (i) obtain a feasible endpoint and a cheap bottleneck bound; (ii) at budget $U$, remove pairs that cannot become edges and force bridges; (iii) compute a conic lower bound over the forced edges; and (iv) branch only on undecided edges. If the root is already proved optimal, the search returns; otherwise a best-first list processes tree sets until the gap is below tolerance or a resource limit is reached. Figure~\ref{fig:shared} shows the two coupled objects.

\subsection{Feasible endpoint and cheap bound}
The Euclidean minimum spanning tree (MST) of the initial positions gives a tree $T_0$ and a feasible conic endpoint. Let $\ell_{\max}$ be its longest edge. Deleting that edge creates a cut $(A,B)$. By the MST cut property, every edge crossing this cut has initial length at least $\ell_{\max}$. Any connected terminal graph crosses the cut with some edge $(i,j)$ satisfying $\|x_i-x_j\|_2\le R$; the triangle inequality gives $\ell_{ij}\le R+2D^\star$. Hence
\begin{equation}\label{eq:bottleneck}
D^\star\ \ge\ L_{\rm mst}=\max(0,(\ell_{\max}-R)/2).
\end{equation}
A bounded tree-exchange pass (swap one tree edge and re-solve) can improve $U$; it never affects the lower bound. The initial MST is not always optimal: in a four-node example it costs 2.074 m, while the full-tree optimum is 1.961 m.

\subsection{Budget geometry and forced edges}
For a feasible $U$, an edge $(i,j)$ can belong to a terminal tree of cost at most $U$ only if
\begin{equation}\label{eq:candidate}
\|p_i-p_j\|_2\le\|x_i-x_j\|_2+\|x_i-p_i\|_2+\|x_j-p_j\|_2\le R+2U.
\end{equation}
Build $H(U)$ on the survivors with those pairs. If $e$ is a bridge of $H(U)$, every connected spanning subgraph of $H(U)$ contains $e$; hence every improving solution contains $e$. Contracting forced edges, deleting forbidden edges, and forcing bridges of the resulting candidate multigraph produces a tree set that contains every potentially improving tree. Parallel edges are retained; contraction is topological only, and all geometric constraints keep the original endpoints.

\subsection{Branch bottleneck bound}
At a search node, let $F$ be the included edges and $Q$ the forbidden ones. We require $F$ to be a forest on the allowed candidate edges. Contracting the components of $F$ and retaining parallel edges gives a multigraph; contraction is topological only, while every original node keeps its own displacement constraint. We then run Kruskal on this multigraph with edge weights $\ell_e=\|p_i-p_j\|_2$, after deleting $Q$ and after forcing the bridges that every completion must contain. The resulting completion is a tree $C$ in the contracted graph and contains $F$; the corresponding original edges connect all survivors.

\begin{lemma}[Branch bottleneck bound]\label{lem:bottleneck}
Let $F$ be a forced forest and let $Q$ be forbidden edges. After contracting $F$, deleting $Q$, forcing all bridges of the resulting multigraph, and retaining parallel edges with their shortest length, let $C$ be the Kruskal completion and $b=\max_{e\in C}\ell_e$. Every completion $T$ consistent with the branch and satisfying $D(T)<U$ has maximum initial edge length at least $b$. Consequently
\begin{equation}\label{eq:branchbn}
D(T)\ge L_{\rm bn}(F,Q)=\max\{0,(b-R)/2\}
\end{equation}
for every such tree $T$.
\end{lemma}
\begin{proof}
Every completion contains $F$ and the forced bridges. If an edge attaining $b$ is forced, it is contained in every completion. Otherwise Kruskal first connects the contracted multigraph at threshold $b$; below $b$ the allowed edges do not connect its components. Any completion must add an edge of length at least $b$. Parallel edges are equivalent to their shortest representative for this threshold, and contraction is only structural. Thus no branch completion has all initial edge lengths below $b$.

For a tree with $D(T)<U$, every edge lies in $H(U)$, so the same argument applies to its completion in the contracted candidate graph. For any feasible endpoint $X$ of $T$ and any edge $(i,j)\in T$, the triangle inequality gives
\[
\ell_{ij}=\|p_i-p_j\|_2\le \|x_i-x_j\|_2+\|x_i-p_i\|_2+\|x_j-p_j\|_2\le R+2D(T).
\]
Because some branch edge has $\ell_e\ge b$, $R+2D(T)\ge\ell_e\ge b$, hence $D(T)\ge(b-R)/2$; taking the non-negative part proves \eqref{eq:branchbn}.
\end{proof}

The lemma is the bound used by the implementation. The completion itself is only a candidate plan: after its fixed-tree SOCP is solved and its endpoint is independently verified, it may update the global upper bound $U$. The bottleneck value and the SOCP endpoint therefore serve different roles, and the algorithm states them separately.

\subsection{Joint mandatory-link bound (JMLB)}
For a forced edge set $F$, orient each $e=(i,j)$ and choose $h_e\in\R^d$. Let
\begin{equation}\label{eq:divergence}
g_i=\sum_{e:a(e)=i}h_e-\sum_{e:b(e)=i}h_e .
\end{equation}
For any feasible $X$ containing $F$,
\begin{align}
\sum_{e=(i,j)}h_e^\top(p_i-p_j)
&=\sum_{e}h_e^\top(x_i-x_j)+\sum_i g_i^\top(p_i-x_i)\notag\\
&\le R\sum_e\|h_e\|_2+\tau\sum_i\|g_i\|_2. \label{eq:dual}
\end{align}
The first equality uses $p_i-p_j=(x_i-x_j)+(p_i-x_i)-(p_j-x_j)$; the inequality uses $\|x_i-x_j\|\le R$ and $\|p_i-x_i\|\le\tau$. Rearranging,
\begin{equation}\label{eq:bound}
\tau\ \ge\ L_F(h)=\frac{\sum_{e}h_e^\top(p_i-p_j)-R\sum_e\|h_e\|_2}{\sum_i\|g_i\|_2},
\end{equation}
with the trivial bound $0$ if the denominator is zero. The tight implemented form is the conic problem
\begin{equation}\label{eq:dualopt}
\max_h\ \sum_{e}h_e^\top(p_i-p_j)-R\sum_e\|h_e\|_2
\quad\text{s.t.}\quad \sum_i\|g_i\|_2\le1,
\end{equation}
whose optimal value is evaluated with the downward margin $\eta=10^{-9}\max(1,R,U,\max_i\|p_i\|_2)$. If $F$ is forced only at budget $B$, the tree set receives $\min(B,L_F(h))$: a solution outside the budget already costs more than $B$. We call this the \emph{Joint Mandatory-Link Bound} (JMLB). The subproblem is a standard SOCP dual; the new element is identifying the mandatory edges and aggregating their shared-endpoint directions into one branch bound. The algorithm uses $\max(L_{\rm bn},L_F)$, not a sum; when $F=\varnothing$, $L_F=0$ and the bottleneck bound remains. JMLB is not a replacement for the per-edge or bottleneck bound. The implemented lower bound is numerical: Clarabel returns the dual vector $h$; the code recenters the coordinates, recomputes the numerator and denominator of \eqref{eq:bound}, returns $0$ when the denominator is non-positive, and subtracts the margin $\eta$. The candidate endpoint is scaled by $1-10^{-11}$, its communication graph is checked, and $U$ is the actual maximum displacement. This is a numerical certificate, not interval arithmetic; the enumeration and overlapping-interval checks are independent cross-checks. Its difference from a per-edge bound is easiest to see at a shared relay. Individually, each incident edge can be satisfied by moving the relay toward that neighbor; those terminal positions are different, but the relay can occupy only one position. The vectors $g_i$ add the incident directions, so the denominator of \eqref{eq:bound} reflects the conflict. A bound that treats the edges independently misses it. For comparison, the per-edge bound of a mandatory set $F$ is $L_{\rm edge}(F)=\max_{e\in F}\max\{0,(\ell_e-R)/2\}$. Since the completion $C$ contains $F$, $L_{\rm bn}\ge L_{\rm edge}(F)$; the bottleneck bound already dominates the independent per-edge bound. JMLB uses the shared-endpoint coupling on the mandatory set $F$; the search takes the maximum with $L_{\rm bn}$, so the two bounds are complementary.

\subsection{Tree-set search and stopping}
A search node stores $(F,Q)$, the forced and forbidden edge decisions. The node represents every spanning tree consistent with those decisions. When a node is popped, its lower bound $l$ is checked first. If $l\ge U-\varepsilon$, the node is pruned and $l$ is recorded; a tree inside it can improve the current plan by at most $\varepsilon$, but its cost is at least $l$. Otherwise the search contracts $F$, removes $Q$, and forces bridges of the resulting candidate multigraph. A disconnected result contains no connected tree and is removed. A Kruskal completion supplies a candidate plan and the branch bottleneck bound \eqref{eq:branchbn}; the dual bound over $F$ is evaluated and the maximum of the inherited, bottleneck, and dual bounds is the new node bound. A candidate completion updates $X$ and $U$ only after its endpoint has been independently checked. If no optional edge remains, the fixed tree is solved and its endpoint is checked: a verified endpoint updates $X$ and $U$. The implementation records the leaf dual bound (or an exact fixed-tree optimum $f(T)$ when separately certified; the implementation reports only the dual bound); if the leaf reaches the pruning threshold it is retired, otherwise its valid bound is recorded and the search continues with other branches; an unresolved leaf is retained in the returned lower bound, not discarded. If the cone solve is unresolved, the leaf bound is recorded and the search continues; dropping it would invalidate the lower bound. If the node or time limit is reached, the node is also returned to the list with its current bound. The remaining node is split on an undecided edge into two disjoint children. After every update, the global lower bound is recomputed from the active and pruned nodes.

\begin{algorithm}[H]
\footnotesize
\caption{Connectivity Restoration Branch-and-Bound (CR-BB)}
\label{alg:search}
\begin{algorithmic}[1]
\Require survivors $p_i$, radius $R$, tolerance $\varepsilon$, node/time limits
\State $T_0\gets$ Euclidean MST; $X\gets$ conic endpoint; $U\gets\max_i\|X_i-p_i\|$
\State apply bounded exchanges; update $U$ if the endpoint improves
\State $B\gets U$; build $H(B)$; $L_0\gets\max(L_{\rm mst},\min(B,L_F(h)))$ on its bridges; $L\gets L_0$
\State search list $\gets\{(F=\varnothing,Q=\varnothing,L)\}$; retired $\gets+\infty$
\While{search list nonempty and $U-L>\varepsilon$ and budget remains}
  \State update $L$ from the current active and pruned bounds
  \State pop the tree set with smallest bound $l$
  \If{$l\ge U-\varepsilon$}
    \State retired $\gets\min(\text{retired},l)$; \textbf{continue}
  \EndIf
  \State contract $F$, remove $Q$, force bridges
  \If{tree set disconnected} \State \textbf{continue} \EndIf
  \State compute Kruskal completion $C$, $L_{\rm bn}(F,Q)$, and dual bound $L_F(h)$
  \State $l\gets\max(l,L_{\rm bn}(F,Q),L_F(h))$
  \If{$l\ge U-\varepsilon$}
    \State retired $\gets\min(\text{retired},l)$; \textbf{continue}
  \EndIf
  \If{node or time limit reached}
    \State push $(F,Q,l)$ with $l$; \textbf{break}
  \EndIf
  \State solve the completion SOCP; verify its endpoint against the communication graph
  \State update $X$ and $U$ if the endpoint is verified and improves $U$
  \If{no optional edge remains}
    \If{the endpoint is verified}
      \State $l\gets\max(l,L_F(h))$
      \If{$l\ge U-\varepsilon$}
        \State retired $\gets\min(\text{retired},l)$; \textbf{continue}
      \Else
        \State retired $\gets\min(\text{retired},l)$; status $\gets$ open bound; \textbf{continue}
      \EndIf
    \Else
      \State record as unresolved (no infeasibility certificate); retired $\gets\min(\text{retired},l)$; \textbf{continue}
    \EndIf
  \EndIf
  \State branch on an undecided edge $e$; push $(F\cup\{e\},Q,l)$ and $(F,Q\cup\{e\},l)$
  \State $L\gets\min(U,\max(L_0,\min(\min_{a\in\mathcal{A}}l_a,\text{retired})))$
\EndWhile
\State $L\gets\min(U,\max(L_0,\min(\min_{a\in\mathcal{A}}l_a,\text{retired})))$
\State \textbf{return} $X$, $[L,U]$, status
\end{algorithmic}
\end{algorithm}

At any stop, let $\mathcal{A}$ be the active tree sets and $\mathcal{R}$ the pruned ones, with valid lower bounds $l_a$ and $l_r$. The returned lower bound is
\begin{equation}\label{eq:globalL}
L=\min\left(U,\max\left(L_0,\min\left(\min_{a\in\mathcal{A}}l_a,\min_{r\in\mathcal{R}}l_r\right)\right)\right),
\end{equation}
with each inner minimum equal to $+\infty$ for an empty set. The two minima are combined before the outer max, so an empty $\mathcal{R}$ cannot erase an active bound; an unresolved leaf enters $\mathcal{R}$ with its retained bound. The next two results make this a global statement.

\begin{proposition}[Branch invariant]\label{prop:invariant}
At every loop point, for every spanning tree $T$ with $D(T)<U$, either $T$ is represented by an active branch or its branch has a recorded valid lower bound $l\le D(T)$.
\end{proposition}
\begin{proof}
Initially the root branch represents every spanning tree. Consider one loop iteration and a tree $T$ with $D(T)<U$ before the iteration.

The candidate filter is safe: every edge of $T$ satisfies $\ell_e\le R+2D(T)<R+2U$, so it is present in $H(U)$. A forced bridge belongs to every connected spanning subgraph of $H(U)$, hence to $T$; a forbidden edge is absent from $T$; contracting $F$ only identifies components already joined by forced edges. If $T$ lies in the popped branch, it corresponds to a completion of the contracted multigraph.

If the branch is pruned at $l\ge U-\varepsilon$, its recorded bound is valid for every tree it contains. Since $T$ is in it, $l\le D(T)$; the branch is recorded. If the branch is disconnected or has no completion, it contains no connected tree, contradicting the existence of $T$. If the search branches on an undecided edge, the include and exclude children are disjoint and exhaustive, so $T$ belongs to exactly one child. If a resource limit is reached, the branch and its bound remain in the active list. Updating $U$ only lowers it; a tree with $D(T)<U_{\rm new}$ also had $D(T)<U_{\rm old}$, and its branch was already represented or recorded. Induction over loop iterations proves the statement.
\end{proof}

\begin{theorem}[Coverage of improving trees]
Every tree that can improve the current upper bound by more than $\varepsilon$ is represented by an active branch or by a recorded branch bound. A tree pruned at threshold $U-\varepsilon$ has cost at least $U-\varepsilon$; a disconnected branch contains no connected tree.
\end{theorem}
\begin{proof}
This is the contrapositive of Proposition~\ref{prop:invariant} for trees with $D(T)<U$: a tree that can improve $U$ by more than $\varepsilon$ has $D(T)<U-\varepsilon<U$, so it is active or recorded and its bound is at most $D(T)$.
\end{proof}

\begin{theorem}[Global guarantee]\label{thm:global}
At any stopping time, \eqref{eq:globalL} satisfies $L\le D^\star\le U$. If the search stops with $U-L\le\varepsilon$, the returned plan has maximum displacement at most $D^\star+\varepsilon$.
\end{theorem}
\begin{proof}
The returned endpoint has cost $U$ and is connected, so $D^\star\le U$. If $D^\star=U$, the truncation in \eqref{eq:globalL} gives $L\le U=D^\star$. Suppose $D^\star<U$. A spanning tree $T_{\rm opt}$ attaining $D^\star$ lies in $H(U)$ and is not removed by a forced/forbidden decision. By Proposition~\ref{prop:invariant}, its branch is active or has a recorded valid lower bound $l\le D^\star$. In either case the branch bound entering the inner minimum of \eqref{eq:globalL} is at most $D^\star$; an empty active or retired set contributes $+\infty$ inside the inner minimum, so it cannot raise the lower bound spuriously. The initial bound satisfies $L_0\le D^\star$, and the truncation by $U$ cannot increase it. Hence $L\le D^\star$. The last statement follows from $U-D^\star\le U-L$.
\end{proof}

In exact arithmetic with exact subproblem solves, the search is complete under the threshold rule: every tree that can improve the incumbent by more than $\varepsilon$ is covered, and every pruned node contributes a valid bound. The returned plan is therefore within $\varepsilon$ of the optimum when the gap closes. In floating point, the implementation uses a downward margin on the conic bound and verifies each endpoint against the communication graph; the displacement interval $[L,U]$ is a numerical guarantee at $\varepsilon$, not a rational proof. Let $K$ be the number of steps used by the verified plan. If $L>(K-1)s$, no plan with $K-1$ steps exists, so $K$ is the minimum step count. This condition is independent of $U-L\le\varepsilon$: it can hold while the displacement interval is still open.
\subsection{Progress-based adaptive stopping (PAS)}
\label{sec:pas-method}
CR-BB can stop at checkpoints before the hard node cap is reached. At a checkpoint, it records the incumbent, the current displacement gap, the number of active tree sets, and the time still available. If the recent gap reduction and projected gain do not justify another stage, the search stops with the incumbent and the bounds at that point. A geometric probe at the checkpoint can raise the global lower bound; this is a bound update, not a branch decision. Because two runs can stop at different nodes, they may return different plans and intervals, and both remain valid. The first check in the main configuration is 64 nodes; the hard cap is 2,000 in the recovery comparison and 2,048 in the budget sweep. At checkpoint $k$, with $g_k=U-L_k$, $\Delta g_k=g_{k-1}-g_k$, $\Delta n_k=n_k-n_{k-1}$, hard node cap $N_{\max}$, deadline $t_{\rm dead}$, and $c_k=\min\{N_{\max}-n_k,(t_{\rm dead}-t_k)\Delta n_k/\max(t_k-t_{k-1},10^{-9})\}$, the stage is promising when $\Delta g_k\ge0.01g_{k-1}$ and $2\Delta g_k c_k/\Delta n_k\ge g_k-\varepsilon$, or when the active queue falls to at most $0.75$ of its previous size; PAS stops after two unpromising stages. The geometric probe bisects $[L,U]$: a disconnected $H(B)$ raises $L$ to $B$, otherwise the forced bridges give $L\leftarrow\max(L,\min(B,L_F))$. Both updates are valid and a failed probe never removes a tree set.

\section{Experiments}
\subsection{Design and metrics}
We use 2,466 UAV performance cases and 31 geometry checks. The performance set contains the $4\times5$ size--loss grid (1,000 cases), the fine $N=1000$ loss scan after removing its overlap with the grid (700), a cross-layout set (300), a correlated-damage set with random, strip, and cross shapes (150), a density set (150), an $N=2000$ set (150), and an independent-layout set (16). It spans 44 distinct layout hashes. The 1,000-case grid is a fixed test set; the recovery comparison reports paired descriptive differences, not population p-values. The 31 geometry checks use their own small radii and are not mixed into the performance tables.

The main configuration is CR-BB with PAS: first check at 64 nodes, hard cap 2,000 nodes, 10 s, and displacement tolerance $\varepsilon=10^{-5}$ m on one CPU thread. The parameter sweep in Section~\ref{sec:adaptive} uses a separate hard cap of 2,048. Solver time is measured from the planner call to its return, excluding replay. SCIP time includes model construction. CR-BB, Shi EA, and Shi OPT receive the same coordinates, survivor identities, radius, movement limit, and replay horizon; R2C uses its published fixed 800-step window, as stated in Appendix~\ref{app:groups}.

A \emph{valid plan} is a connected terminal configuration whose moves have been replayed. \emph{Continuous gap attainment} means $0\le U-L\le\varepsilon$. \emph{First-connectivity steps} are the number of one-metre moves at which the replayed network first becomes connected. Across the 4,932 saved-plan replays of the 2,000-node cap and CR-BB with PAS runs, this count equals the plan step count in every run. All experiments use one-metre steps, $s=1$ m. \emph{Integer optimality} is a separate statement: if the verified plan uses $K$ steps and $L>(K-1)s$, no plan with $K-1$ steps exists. The three outcomes are reported separately.

\subsection{Recovery performance}
\begin{figure}[t]
\centering
\includegraphics[width=.98\linewidth]{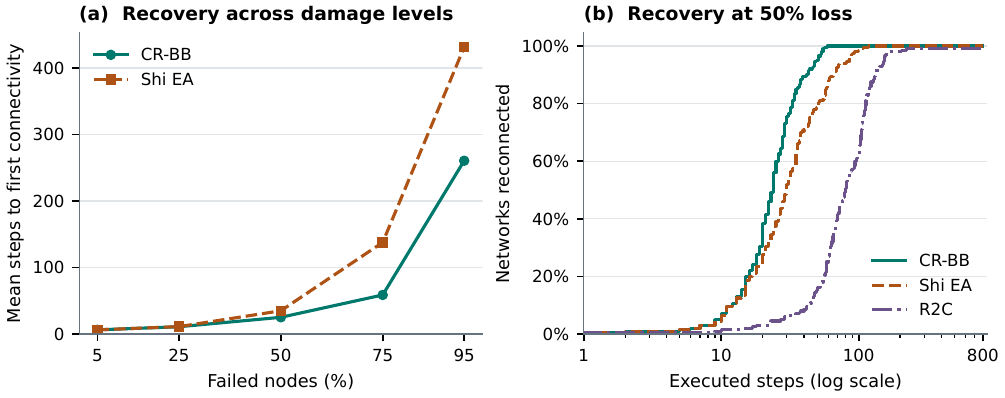}
\caption{Recovery steps of CR-BB with PAS and Shi EA on matched failure cases. (a) Each point averages 200 cases at one loss level, with 50 cases from each of four initial network sizes. (b) Fraction of the 200 networks at 50\% loss that have regained connectivity by each step. R2C uses its own fixed 800-step window and coverage objective; the curve is an application reference, not a like-for-like optimality comparison. Two R2C runs remained disconnected at that horizon.}
\label{fig:recovery}
\end{figure}

Table~\ref{tab:recovery} and Figure~\ref{fig:recovery} compare the same failure scenarios. On the 1,000-case grid, the replayed CR-BB with PAS plans first connected in 72.35 steps on average, compared with 124.32 for Shi EA, a reduction of 41.8\% in the mean. On the common 50\% subset of 200 cases, the means are 25.13 and 34.97 steps. Shi EA is the deterministic $k=1$, zero-minimum-separation specialization of the edge-augmentation baseline with SOCP refinement: it constructs augmented links, solves the fixed-tree SOCP, and uses no random seed, training, or internal budget; each case is one run. The same connectivity check and one-metre replay are applied to its endpoint. Appendix~\ref{app:groups} reports the R2C application reference: 198 of the 200 networks recover by the 800-step horizon, with a capped mean of 91.5 steps and 84.35 steps over the 198 recovered networks.

\begin{table}[t]
\centering
\small
\caption{Paired recovery on identical inputs. Entries are mean first-connectivity steps from replayed plans; CR-BB uses PAS. The grid and the 50\% subset are paired against Shi EA; reduction is $100(1-\text{CR-BB}/\text{Shi EA})$. R2C is an application reference in Appendix~\ref{app:groups}.}
\label{tab:recovery}
\begin{tabular}{lrrrr}
\toprule
Cohort & Cases & CR-BB & Shi EA & Reduction in mean (\%) \\
\midrule
Grid & 1,000 & 72.35 & 124.32 & 41.8 \\
50\% loss & 200 & 25.13 & 34.97 & 28.1 \\
\bottomrule
\end{tabular}

\end{table}

\subsection{Solver accuracy and computation}
Table~\ref{tab:solver} uses the same 60-case subset for CR-BB, the spanning-tree MISOCP solved with SCIP, and Shi OPT. The subset is the fixed stratified sample from the 1,000-case grid: four network sizes by five damage levels, taking the first three archived cases in each cell by case number. All 60 belong to the 2,466-case performance set.

CR-BB and the MISOCP each returned a connected verified endpoint on all 60 assigned cases. They met $0\le U-L\le10^{-5}$ m in 54 cases (90.0\%) and 33 cases (55.0\%), respectively. Shi OPT could build and run on 8 cases because of the 100,000-binary-variable guard; it returned a valid endpoint on all 8 and met the gap on 3 (37.5\% of the runs, 5.0\% of the 60 assigned cases). The guard protects memory; instances outside it are left open rather than declared unsolvable.

The times are from a same-machine, per-case paired rerun with one Basic Linear Algebra Subprograms (BLAS) thread per worker and a fixed method order (CR-BB, then the MISOCP); each uses its own 10 s internal search limit. A no-PAS CR-BB control also certified 54/60 (median 0.055 s); CR-BB with and without PAS have the same 54 paired certifications. On the eight Shi OPT cases, CR-BB certifies all 8 and the MISOCP 3. The CR-BB/MISOCP 2×2 is 33 both, 21 CR-BB-only, 0 MISOCP-only, and 6 neither; under the $|U-L|\le\varepsilon$ convention the MISOCP count is 36/60 rather than 33/60; the three negative-gap records lie in the CR-BB-only cell, so that cell becomes 18 and both becomes 36. Table~\ref{tab:solver} reports total elapsed time from the solver call; SCIP build time is included. The median total times are 0.041 s for CR-BB, 6.468 s for the MISOCP, and 10.301 s over the eight Shi OPT runs. These are paired only on the instances where each method ran; they are descriptive paired comparisons on one machine.

Gap counts use the strict criterion $0\le U-L$. Three MISOCP records have a small negative $U-L$ (magnitude at most $1.8\times10^{-11}$ m); they are retained as numerical inconsistencies and are not counted as meeting the gap. No negative gap occurs in the rerun CR-BB records. On the 33 instances where both CR-BB and the MISOCP meet the gap, all intervals intersect; the largest lower-bound difference is $9.30\times10^{-7}$ m.

\begin{table}[t]
\centering
\small
\caption{Results on the same 60 instances with a 10 s internal solver search limit. ``Run'' counts instances that reached the solver; all returned endpoints were independently verified. ``Gap met'' gives the count over all 60 assigned cases and over the runs that started. Times are total elapsed times; SCIP build/initialization and solver overhead are included, which is why the Shi median can slightly exceed the 10 s optimization limit. CR-BB uses the 2,000-node cap. Shi OPT stopped at the binary-variable guard on 52 cases.}
\label{tab:solver}
\begin{tabular}{lrrrrr}
\toprule
Method & Assigned & Run & \multicolumn{2}{c}{Gap met} & Median time (s) \\
\cmidrule(lr){4-5}
 & & & assigned & run & \\
\midrule
CR-BB with PAS & 60 & 60 & 54 (90.0\%) & 54 (90.0\%) & 0.041 \\
Spanning-tree MISOCP & 60 & 60 & 33 (55.0\%) & 33 (55.0\%) & 6.468 \\
CR-BB without PAS & 60 & 60 & 54 (90.0\%) & 54 (90.0\%) & 0.055 \\
Shi OPT (SCIP) & 60 & 8 & 3 (5.0\%) & 3 (37.5\%) & 10.301 \\
\bottomrule
\end{tabular}

\end{table}

\subsection{Branch JMLB at search nodes}
We evaluate JMLB end to end on the 200 cases at 50\% loss. The recovery comparison uses a 2,000-node cap; the 16-setting sweep and this paired ablation use 2,048 for both arms. Both arms share the same root JMLB bound, initial witness, candidate threshold, branch order, tolerance $10^{-5}$ m, and 10 s budget; only the branch JMLB is switched. The off arm retains candidate filtering, forced bridges, the Kruskal bottleneck bound, and leaf dual bounds, so the comparison isolates the branch JMLB.

JMLB meets the prescribed gap on 200/200 cases (100.0\%); the off arm meets it on 182/200 (91.0\%). The paired outcomes are 182 both, 18 JMLB-only, 0 off-only, and 0 neither. Mean expanded nodes fall from 49.6 to 0.7, and total solver time from 196.8 to 15.2 s. The 18 additional attainments are cases in which the branch bound reaches $U-\varepsilon$ before the search would otherwise exhaust its budget. Table~\ref{tab:jmlb200} summarizes this run. Of the 200 cases, 181 are already closed at the root; 19 enter tree search. The unconditional mean is dominated by 18 time-limited off runs, so both arms have median 0 expanded nodes; among the 19 branched cases the mean is 522.2 versus 7.1 nodes and the median is 575 versus 5. The off/on median wall times are 0.048 s and 0.048 s; the mean is dominated by the time-limited tail.

To separate the bound from the end-to-end effect, we collected 456 branch snapshots from the 19 cases that entered tree search, taking 24 snapshots per case; these snapshots come from searches in which the branch JMLB was disabled. On each snapshot we evaluate three quantities with the same $U$: the per-edge bound $L_{\rm edge}(F)$, the Kruskal bottleneck bound $L_{\rm bn}(F,Q)$, and JMLB $L_F(h)$ using the same conic dual. The per-edge bound is never larger than the bottleneck bound in these snapshots, consistent with $C\supseteq F$. JMLB improves on $\max(L_{\rm edge},L_{\rm bn})$ in 418 of 456 snapshots (91.7\%), with a median lift of 2.35 m and a maximum of 20.74 m. It crosses the pruning threshold in 421 snapshots (92.3\%), compared with 3 snapshots for the baseline bounds. The snapshots are the first 24 distinct forced sets encountered in each off search; all 19 cases contribute 24. They are a diagnostic sample rather than a uniform sample of the 200 cases. The per-edge bound is dominated by the bottleneck bound; the substantive comparison is JMLB against $L_{\rm bn}$.

\begin{figure}[t]
\centering
\includegraphics[width=.98\linewidth]{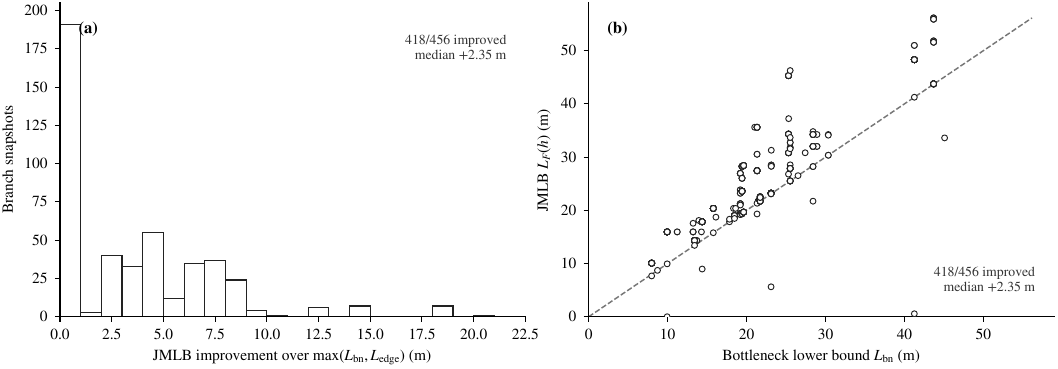}
\caption{Branch JMLB on the 200 matched 50\%-loss cases. (a) Distribution of the JMLB lift over $\max(L_{\rm edge},L_{\rm bn})$ on 456 shared branch snapshots from 19 cases (24 each); 418 are positive and the median is 2.35 m; the first bin includes 38 zero-lift snapshots. (b) Bottleneck bound against JMLB on the same snapshots; points above the diagonal are branches that JMLB tightens. The end-to-end paired result is 182/200 gap attainments without branch JMLB and 200/200 with it.}
\label{fig:jmlb}
\end{figure}

\subsection{Progress-based adaptive stopping}
\label{sec:pas-results}

\label{sec:adaptive}
We sweep the fixed node cap and the first PAS check over powers of two from 16 to 2,048 nodes. Each setting is tested on all 2,466 cases. The PAS configurations share a hard cap of 2,048 nodes; the main recovery comparison uses the separate 2,000-node cap. Figure~\ref{fig:pas} and Table~\ref{tab:scan} in Appendix~\ref{app:scan} report the full curves and the exact configuration statistics.

At a first check of 64, CR-BB with PAS and CR-BB with a fixed 2,048-node cap both met the prescribed gap in 92.0\% of runs. PAS reduced cumulative solver time from 2,849.2 to 2,254.1 s (20.9\%) and mean expanded nodes from 61 to 42 (31.1\%). The plans were 0.36 steps longer on average. Of the 2,466 cases, 2,268 met the gap in both configurations with identical plan steps; 196 met it in neither and account for 591.06 s of the 595.05 s net saving; two met it in one configuration only. In the 196 cases where neither met it, PAS returned longer plans in 60 cases and shorter plans in two; among the 60 longer plans the median increase was 10 steps and the maximum 50.

The fixed-512 setting has approximately the same mean number of expanded nodes, but meets the gap in 2,265 cases rather than 2,269 and takes 2,433.6 s rather than 2,254.1 s. The fixed-256 setting takes 1,836.0 s and meets 2,254 cases; it is a lower-time point with a larger plan cost and fewer attainments. The complete time-quality frontier is shown in Figure~\ref{fig:pas}; the 64-node check is the earliest setting that matches the 2,048-node fixed cap, an operating point on this frontier. The setting was selected on this sweep and is treated as an operating point.

The full 2,000-node-cap archive also isolates the geometric probe. PAS with the probe meets 2,272 cases in 2,255.9 s; with the probe disabled it meets 2,270 in 2,273.1 s; the 2,000-node cap meets 2,270 in 2,855.4 s. The probe changes a bound in 12 cases and adds two gap attainments (2,270 to 2,272). The one-case difference between the 2,048- and 2,000-node fixed runs is an archive effect; both are labeled in Appendix~\ref{app:scan}. The main time saving survives without it (20.4\% versus the fixed budget). PAS is therefore primarily a stopping rule; the probe is a secondary tightening mechanism. Table~\ref{tab:probe} lists the three runs.

\begin{figure}[t]
\centering
\includegraphics[width=.98\linewidth]{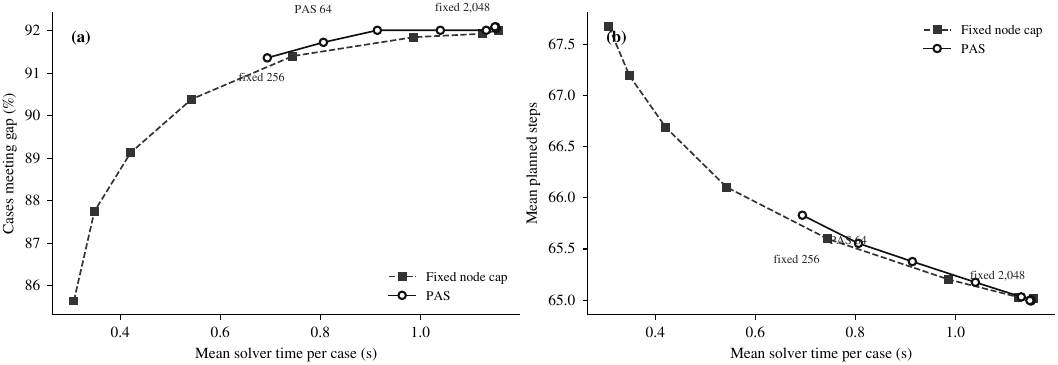}
\caption{Time--quality trade-offs of the 16 budget settings on 2,466 cases. The horizontal axis is mean solver time per case. (a) Fraction meeting $U-L\le10^{-5}$ m. (b) Mean planned steps. Squares: fixed node caps; circles: PAS with a 2,048-node hard cap. Key operating points are marked; the vertical axes are truncated to the plotted ranges. The full configuration table is Table~\ref{tab:scan}.}
\label{fig:pas}
\end{figure}

\subsection{A strip-loss case}
\begin{figure}[t]
\centering
\includegraphics[width=.68\linewidth]{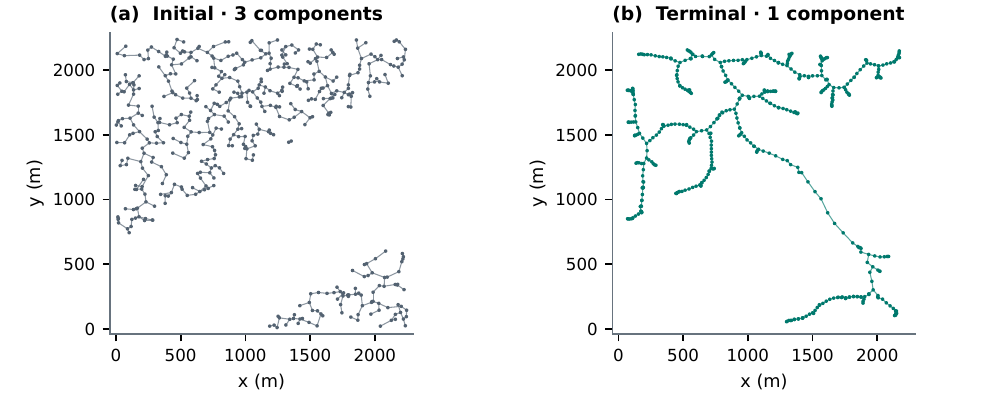}
\caption{Connectivity restoration after strip damage in a 1,000-node network with 500 survivors. Each panel shows the same aircraft at the same spatial scale; edges form the connected terminal communication graph, projected onto the horizontal plane; $R=120$ m. The network first regained connectivity at step 494.}
\label{fig:case}
\end{figure}

Figure~\ref{fig:case} shows a verified plan for a strip-loss case with 1,000 initial nodes and 500 survivors. The initial communication graph has three components; the terminal graph has one. The replayed plan first connects at step 494. The returned lower bound is 493.37623 m; with a one-metre step, 493 steps can cover at most 493 m, so 494 steps is the minimum. This is the case-level version of the displacement-bound guarantee: the plan is executable, and the bound excludes every shorter integer plan.

\section{Discussion and Conclusion}
We have formulated minimum displacement as a global optimization problem with two outputs: a terminal configuration and bounds on its optimal value; dividing the bounds by the common speed gives the corresponding time interval. For a fixed tree, the best terminal positions solve a second-order cone program. The harder step is choosing the tree and deciding when further search is worthwhile. JMLB addresses the first part for a set of trees at once: a relay shared by several required links has one terminal position, so incident directions must be evaluated together. The bound removes a tree set only after accounting for that coupling.

The experiments distinguish three outcomes that are often conflated. Continuous gap attainment is $U-L\le10^{-5}$ m. Integer optimality is $L>(K-1)s$ for a verified $K$-step plan. First-connectivity steps come from replaying the movement. The strip case makes the distinction concrete: the lower bound exceeds 493 m, so a 493-step plan cannot connect, while the replayed 494-step plan does. The integer certificate can also hold while the displacement interval remains open, as in a 1,000-node, 850-failure case, where $L=99.8976$ m proves the returned 100-step plan optimal even though $U-L$ is still $0.042$ m.

PAS changes when the search stops, and therefore which incumbent and bounds are returned. It saves time mainly when neither configuration meets the prescribed gap: on the 2,466-case sweep, the 196 cases where neither met it account for 591.06 s of the 595.05 s net saving. When both configurations meet the gap, the plans have identical steps. The 64-node first check is the selected operating point because it reaches the 92.0\% gap-attainment plateau; later checks buy at most two more gap attainments at several hundred seconds. A 223-step certificate case illustrates the cost: PAS stops at 256 nodes while the fixed 2,000-node cap proves optimality at 513 nodes, yet both return the same plan; the difference is the certificate.

Within this setting, CR-BB returns a plan and displacement bounds, JMLB supplies the global comparison, and PAS, a budget-control policy, reduces computation while preserving the bounds. Extending the bounds to obstacles, collision avoidance, and rolling-horizon replanning is future work.

\appendix
\section{Scenario sets}\label{app:scenarios}
Table~\ref{tab:scenarios} summarizes the fixed performance set. All performance cases are three-dimensional with $R=120$ m; the 31 enumeration checks in Appendix~\ref{app:correct} use their own radii. The sets differ in size, damage fraction, layout, and damage shape. The 44 layout hashes are parent deployments, not independent samples: multiple damage masks share a layout, so inferential statements are made at the layout level where appropriate. The replay horizon is $H_{\rm rep}=600,800,1280,1800,2560$ for $N=100,200,500,1000,2000$. Damage is random in 2,366 cases, strip in 50, and cross in 50; the 1,000-case grid uses four parents with 250 masks each, the fine scan one parent with 700, and the remaining sets 2--150 masks per parent; over all 44 parents the range is 2--995.

\begin{table}[H]
\centering
\small
\caption{Performance sets. Each row is a fixed test set; the total is the 2,466-case cohort used in the main text. Sizes and loss ranges are minimum--maximum within the row.}
\label{tab:scenarios}
\begin{tabular}{lrrrr}
\toprule
Set & Cases & Sizes & Loss range & Layout hashes$^\ast$ \\
\midrule
A: size--loss grid & 1,000 & 100--1,000 & 5--95\% & 4 \\
B: fine loss scan & 700 & 1,000 & 10--90\% & 1 \\
C: cross-layout & 300 & 100--1,000 & 25--75\% & 20 \\
D: correlated damage & 150 & 1,000 & 50\% & 5 \\
Density & 150 & 1,000 & 50\% & 15 \\
$N=2000$ & 150 & 2,000 & 25--75\% & 1 \\
Independent layouts & 16 & 100--1,000 & 50\% & 8 \\
\midrule
Total & 2,466 & 100--2,000 & 5--95\% & 44 distinct \\
\bottomrule
\end{tabular}

\end{table}

{\footnotesize $^\ast$Layout hashes overlap across sets; 44 is the number of distinct parent deployments, not the column sum.}

\section{Grouped results and the R2C reference}\label{app:groups}
Table~\ref{tab:groups} reports the seven performance sets after the overlap between sets A and B has been removed. ``Gap met'' and ``integer'' use the CR-BB with PAS terminal configurations; the recovery means use the replayed first-connectivity steps. The fine-scan row has 700 rather than 950 cases because the 250 shared with the grid are counted once. The median solver time is from the same CR-BB with PAS run.

\begin{table}[H]
\centering
\scriptsize
\caption{Performance by set after deduplication from the 2,000-node-cap replay archive. Loss is the range of failed-node fractions. CR-BB with PAS and Shi EA are mean first-connectivity steps. ``Gap met'' is the fraction with $U-L\le10^{-5}$ m; ``integer'' is the fraction with a verified plan satisfying $L>(K-1)s$. Times are medians over the set.}
\label{tab:groups}
\begin{tabular}{lrrrrrrr}
\toprule
Set & Cases & Loss & CR-BB & Shi EA & Gap met (\%) & Integer (\%) & Median time (s) \\
\midrule
Grid & 1,000 & 5--95\% & 72.35 & 124.32 & 89.6 & 89.5 & 0.085 \\
Fine scan & 700 & 10--90\% & 47.72 & 124.87 & 88.0 & 88.3 & 0.229 \\
Cross-layout & 300 & 25--75\% & 32.90 & 62.94 & 99.7 & 99.7 & 0.052 \\
Correlated damage & 150 & 50\% & 240.67 & 294.00 & 99.3 & 99.3 & 0.103 \\
Density & 150 & 50\% & 21.81 & 30.07 & 100.0 & 100.0 & 0.110 \\
$N=2000$ & 150 & 25--75\% & 38.60 & 100.48 & 97.3 & 97.3 & 0.373 \\
Independent layouts & 16 & 50\% & 27.75 & 40.62 & 100.0 & 100.0 & 0.048 \\
\bottomrule
\end{tabular}

\end{table}

R2C is retained as an application reference, not as a second optimality baseline: its published objective also includes coverage, and its 800-step execution window is fixed, while the replay horizon is $H_{\rm rep}=600,800,1280,1800,2560$ for $N=100,200,500,1000,2000$; R2C is therefore censored earlier than the replay for the two largest sizes of the 200-case cohort. On the 200 shared 50\%-loss cases it recovers 198 networks by step 800; the capped mean is 91.5 steps with both censored runs in the denominator, and 84.35 steps over the 198 recovered networks. The two censored runs remain in the cumulative curve of Figure~\ref{fig:recovery}.

\begin{figure}[t]
\centering
\includegraphics[width=.98\linewidth]{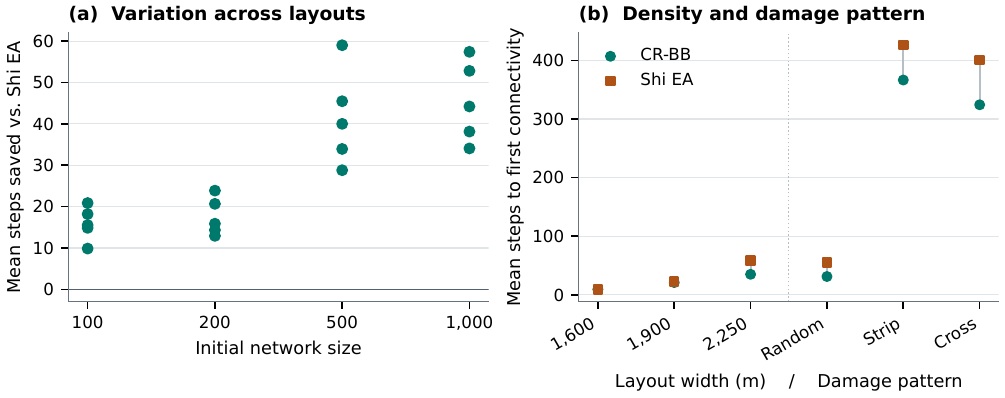}
\caption{Recovery across layouts, deployment widths, and damage patterns. Each point in (a) is the mean paired difference over 15 masks on one layout; positive values favor CR-BB with PAS. In (b), each pair of markers is the mean over 50 scenarios with 50\% loss and 1,000 initial nodes.}
\label{fig:layouts}
\end{figure}

\section{Budget sweep and probe}\label{app:scan}
Table~\ref{tab:scan} lists all 16 configurations from the 2,466-case sweep. The PAS configurations use a 2,048-node hard cap and the stated first check; the fixed configurations use the stated hard cap. Each cell is one run on every case, 39,456 calls in total. Table~\ref{tab:probe} isolates the geometric probe in the separate 2,000-node-cap archive.

\begin{table}[H]
\centering
\scriptsize
\setlength{\tabcolsep}{3.5pt}
\caption{Complete fixed and PAS budget sweep from the 2,048-node archive. ``Gap met'' counts $U-L\le10^{-5}$ m. Total time is the sum over 2,466 cases; the PAS rows use a first check at the stated node count.}
\label{tab:scan}
\begin{tabular}{lrrrrr}
\toprule
Configuration & Gap met & Mean steps & Mean time (s) & Total time (s) & Mean nodes \\
\midrule
Fixed 16 & 2,112 & 67.67 & 0.307 & 758 & 3 \\
Fixed 32 & 2,164 & 67.20 & 0.348 & 859 & 5 \\
Fixed 64 & 2,198 & 66.69 & 0.421 & 1037 & 9 \\
Fixed 128 & 2,229 & 66.11 & 0.543 & 1338 & 15 \\
Fixed 256 & 2,254 & 65.60 & 0.745 & 1836 & 26 \\
Fixed 512 & 2,265 & 65.20 & 0.987 & 2434 & 42 \\
Fixed 1024 & 2,267 & 65.03 & 1.125 & 2775 & 57 \\
Fixed 2048 & 2,269 & 65.02 & 1.155 & 2849 & 61 \\
PAS 16 & 2,253 & 65.83 & 0.694 & 1711 & 29 \\
PAS 32 & 2,262 & 65.56 & 0.806 & 1988 & 35 \\
PAS 64 & 2,269 & 65.38 & 0.914 & 2254 & 42 \\
PAS 128 & 2,269 & 65.17 & 1.040 & 2565 & 50 \\
PAS 256 & 2,269 & 65.03 & 1.132 & 2791 & 59 \\
PAS 512 & 2,271 & 65.00 & 1.151 & 2838 & 61 \\
PAS 1024 & 2,270 & 64.99 & 1.151 & 2839 & 61 \\
PAS 2048 & 2,271 & 65.00 & 1.149 & 2833 & 61 \\
\bottomrule
\end{tabular}

\end{table}

\begin{table}[H]
\centering
\small
\caption{Geometric-probe ablation in the 2,000-node-cap archive. ``Gap met'' counts $U-L\le10^{-5}$ m. The probe changes only a small number of cases; the main stopping-rule saving remains without it.}
\label{tab:probe}
\begin{tabular}{lrrr}
\toprule
Configuration & Gap met & Total time (s) & Mean steps \\
\midrule
Fixed 2,000 & 2,270 & 2,855.4 & 65.00 \\
PAS 64, probe off & 2,270 & 2,273.1 & 65.37 \\
PAS 64, probe on & 2,272 & 2,255.9 & 65.38 \\
\bottomrule
\end{tabular}

\end{table}

\section{JMLB ablation}\label{app:jmlb}
Table~\ref{tab:jmlb200} gives the end-to-end result on the 200 cases at 50\% loss. Both arms share the root bound, initial witness, candidate threshold, branch order, tolerance, 2,048-node cap, and 10 s budget. The off arm still retains the Kruskal bottleneck bound, forced bridges, candidate filtering, and leaf duals. Table~\ref{tab:abl19} preserves the original 19-case, 500-node-cap two-by-two experiment as a separate, selected-case record; it is not used to estimate the population effect.

\begin{table}[H]
\centering
\small
\caption{End-to-end JMLB ablation on the 200 matched 50\%-loss cases. ``Gap met'' is $U-L\le10^{-5}$ m; times are sums over the 200 cases.}
\label{tab:jmlb200}
\begin{tabular}{lrr}
\toprule
Quantity & JMLB off & JMLB on \\
\midrule
Cases meeting \(10^{-5}\) m gap & 182/200 & 200/200 \\
Mean expanded nodes & 49.6 & 0.7 \\
Median expanded nodes & 0 & 0 \\
Root-closed cases & 181 & 181 \\
Time-limited cases & 18 & 0 \\
Total solver time (s) & 196.8 & 15.2 \\
\bottomrule
\end{tabular}
\par\smallskip
{\footnotesize Paired outcomes: 182 both, 18 on only, 0 off only, 0 neither.\\
Branch snapshots: 456 (first 24 distinct forced sets per case, 19 cases); JMLB improves 418 (91.7\%); crosses the pruning threshold in 421 (92.3\%) versus 3 without it; median lift 2.35 m, maximum 20.74 m.}

\end{table}

\begin{table}[H]
\centering
\small
\caption{Original two-by-two bound ablation on the 19 cases that required branching under the root bound; 500-node cap and 10 s. ``Root bound'' is evaluated at the root; ``branch bound'' is the JMLB evaluation inside a search node. This is a selected-case diagnostic, not the population estimate.}
\label{tab:abl19}
\begin{tabular}{llrrr}
\toprule
Root bound & Branch bound & Gap met (n) & Median nodes & Median search (s) \\
\midrule
On & On & 19 & 5 & 0.066 \\
On & Off & 1 & 500 & 4.217 \\
Off & On & 19 & 5 & 0.060 \\
Off & Off & 1 & 500 & 4.393 \\
\bottomrule
\end{tabular}

\end{table}

\section{Correctness and the integer boundary}\label{app:correct}
Table~\ref{tab:correct} reports the 31 full-enumeration geometry checks. The 30 random two- and three-dimensional cases use $R=30$ m and the seven-node shared-relay case uses $R=1.5$ m; the largest enumeration is 16,807 trees and the maximum interval deviation is $6.8\times10^{-8}$ m.

A separate 200-node integer-boundary case is checked, not one of the 31 rows. The stored floating endpoint replays in 10 steps; an independently stored exact plan uses 9 steps. In the exact plan the maximum displacement is 9 m and a tree edge is exactly 120 m. At an eight-step budget every allowable edge has length at most $R+2(8)=136$ m, while the shortest crossing edge of the relevant cut is 138 m; every eight-step plan therefore leaves that cut disconnected. The 10-step value is the replay of the stored floating endpoint, not a second model optimum. The continuous record is $L=8.9999999862$, $U=9.0000000011$ moves, gap $1.49\times10^{-8}$; the exact witness gives $K^\star=9$, so the 10-step endpoint is not integer-optimal. The cut has 94 nodes on one side and 6 on the other; its shortest crossing edge is 138 m at the initial endpoints $(974,302,0)$ and $(992,410,84)$.

\begin{table}[H]
\centering
\small
\caption{Full-enumeration correctness checks. These 31 inputs are not part of the 2,466 performance instances.}
\label{tab:correct}
\begin{tabular}{lrrrr}
\toprule
Check & Cases & Passed (\%) & Max deviation (m) & Max trees/example \\
\midrule
2-D/3-D enumeration, $n=4,5,6$ & 30 & 100.0 & 6.8e-08 & 1,296 \\
Shared relay, $n=7$ & 1 & 100.0 & 1.3e-10 & 16,807 \\
\bottomrule
\end{tabular}

\end{table}

The integer certificate is independent of the continuous gap. In the full sweep, a 1,000-node, 850-failure case remains open with $L=99.8976$ m and $U=99.9401$ m, but $L>99$ proves that the returned 100-step plan is integer-optimal. Conversely, a 223-step case met the prescribed gap under the 2,000-node cap cap at 513 nodes and was stopped early by PAS at 256 nodes; both return a 223-step plan. The difference is the width of the certified interval.

\section{MISOCP formulation}\label{app:mip}
The spanning-tree MISOCP uses a binary $y_e$ for each candidate edge $e=(i,j)$, continuous displacements $u_i\in\R^d$, a displacement variable $\tau$, and a single-commodity flow $f_{ij}$ on every oriented candidate arc. It imposes
\begin{align}
&\sum_e y_e=n-1,\qquad
\sum_{j:(i,j)}f_{ij}-\sum_{j:(j,i)}f_{ji}=n-1 \text{ at the root},\quad -1 \text{ elsewhere},\\
&0\le f_{ij}\le (n-1)y_e,\qquad
\|u_i\|_2\le\tau,\\
&\|p_i-p_j+u_i-u_j\|_2\le M_{ij}-(M_{ij}-R)y_e,
\end{align}
with $M_{ij}=\max\{R,\ell_{ij}+2U_0\}+\varepsilon$, where $U_0$ is an independent feasible upper bound. The first two lines select exactly $n-1$ edges and enforce connectivity by flow; the last line is a pair-specific big-M relaxation of the terminal edge constraint. The model is solved with SCIP 9 through PySCIPOpt 5.2.1, with one thread, $10^{-5}$ m absolute gap target, and $10$ s optimization limit. The independent warm start is the same feasible tree used by CR-BB. Shi OPT is the $k=1$, zero-minimum-separation mixed-integer quadratically constrained program (MIQCP) specialization; its binary count is much larger because the reverse-distance implication is retained. The model implementation is included in the reproduction package.

\section{Reproduction and data availability}
The reproduction package is available from the corresponding author upon request. It contains the case manifests, selected identifiers, per-case result tables, saved terminal configurations, independent connectivity and displacement checks, and the scripts that regenerate the figures. PySCIPOpt 5.2.1 (SCIP 9) is used for the MISOCP rerun. All timings use one basic linear algebra (BLAS) thread per worker; the exact rerun uses four workers, and the JMLB experiment uses twelve. The reproduction script is the entry point for the packaged cases.

\bibliographystyle{unsrtnat}
\bibliography{refs}
\end{document}